\documentclass[twocolumn,showpacs,preprintnumbers,amsmath,amssymb,prb,superscriptaddress]{revtex4}

\usepackage{graphicx}
\usepackage{dcolumn}
\usepackage{bm}

\begin{document}

\title{Tunable Fano effect in parallel-coupled double quantum dot system}
\author{Haizhou Lu} \email{luhz@castu.tsinghua.edu.cn}
\affiliation{Center for Advanced Study, Tsinghua University,
Beijing 100084, P.R.China}
\author{Rong L\"{u}}
\affiliation{Center for Advanced Study, Tsinghua University,
Beijing 100084, P.R.China}
\author{Bang-Fen Zhu}\email{bfzhu@castu.tsinghua.edu.cn}
\affiliation{Center for Advanced Study, Tsinghua University,
Beijing 100084, P.R.China} \affiliation{Department of Physics,
Tsinghua University, Beijing 100084, P.R.China}
\date{\today}

\begin{abstract}
With the help of the Green function technique and the equation of
motion approach, the electronic transport through a
parallel-coupled double quantum dot(DQD) is theoretically studied.
Owing to the inter-dot coupling, the bonding and antibonding
states of the artificial quantum-dot-molecule may constitute an
appropriate basis set. Based on this picture, the Fano
interference in the conductance spectra of the DQD system is
readily explained. The possibility of manipulating the Fano
lineshape in the tunneling spectra of the DQD system is explored
by tuning the dot-lead coupling, the inter-dot coupling, the
magnetic flux threading the ring connecting dots and leads, and
the flux difference between two sub-rings. It has been found that
by making use of various tuning, the direction of the asymmetric
tail of Fano lineshape may be flipped by external fields, and the
continuous conductance spectra may be magnetically manipulated
with lineshape retained. More importantly, by adjusting the
magnetic flux, the function of two molecular states can be
exchanged, giving rise to a swap effect, which might play a role
as a qubit in the quantum computation.
\end{abstract}

\maketitle

\section{Introduction}
When the phase coherence of electrons passing through a mesoscopic
system is retained, a number of quantum interference phenomena
will occur. Recent advances in nano-technologies have attracted
much attention to the quantum coherence phenomena in the resonant
tunneling processes of the quantum dot(QD) systems,\cite{nato} in
which the typical length scale can be shorter than or comparable
to the mean free path of electrons and the wave nature of
electrons plays a decisive role. In the past decade, the widely
adopted method of studying the phase coherence of a traversing
electron through a QD is to measure the magnetic-flux-dependent
current through an Aharonov-Bohm(AB) interferometer by inserting
the QD into one of its arms.\cite{ab} The observed
magnetic-oscillation of the current will be the indication of the
coherent transport through the QD, provided that at least partial
coherence of electrons is kept.
\cite{Holleitner2001,Holleitner2002,Holleitner2003,Blick2003,yacoby,schuster,buks,wiel,yji,Kobayashi2002,Kobayashi2003}

Fano resonances is another good probe for the phase coherence in
the QD
system.\cite{Kobayashi2002,Kobayashi2003,gores,zacharia,johnson}
It is known that the Fano resonance stems from quantum
interference between resonant and nonresonant
processes,\cite{fano} and manifests itself in spectra as
asymmetric lineshape in a large variety of experiments. Unlike the
conventional Fano
resonance,\cite{conventional1,conventional2,conventional3,conventional4}
the Fano effect in QD system has its advantage in that its key
parameters can be readily tuned. Suppose that a discrete level
inside the QD acts as a Breit-Wigner-type scatter and is broadened
by a factor of $\Gamma$ due to couplings with the continua in
leads. The key to realize the Fano effect in the conductance
spectra is that, within $\Gamma$ the phase of the electron should
smoothly change by $\pi$ on the resonance.\cite{schuster}  The
first observation of the Fano lineshape in the QD system was
reported by G\"{o}res \emph{et al}.\cite{gores,zacharia} in
 the single-electron-transistor experiments. Recently, K.
Kobayashi \emph{et al.} carried out research on magnetically and
electrostatically tunable Fano effect in a QD embedded in an AB
ring,\cite{Kobayashi2002,Kobayashi2003} and A.C. Johnson \emph{et
al}. investigated a tunable Fano interferometer consisting of a QD
coupled to a one-dimensional channel via tunneling and observed
the Coulomb-modified Fano resonances.\cite{johnson}

The double quantum dot(DQD) system, including the series-coupled
DQD\cite{Waugh1995,Blick1998,Oosterkamp1998,Qin2001} and
parallel-coupled
DQD,\cite{Holleitner2001,Holleitner2002,Holleitner2003,Blick2003,jcchen}
makes the quantum transport phenomena rich and varied. The
parallel-coupled DQD system is of particular interest, in which
two QD's are respectively embedded into opposite arms of the AB
ring, coupled each other via barrier tunneling, and coupled to two
leads roughly equally. As a controllable two-level system, it is
appealing for the parallel-coupled DQD system to become one of
promising candidates for the quantum bit in quantum computation
based on solid state devices.\cite{loss,hu} The entangled quantum
states required for performing the quantum computation demand a
high degree of phase coherence in the system.\cite{highcoherence}
Being a probe of phase coherence,\cite{clerk} if the Fano effect
in the parallel-coupled DQD system is tunable and exhibits the
swap effect, it is certainly of practical importance.

Inspired by recent experimental advances in the parallel
DQD,\cite{Holleitner2001,Holleitner2002,Holleitner2003,Blick2003,jcchen}
several groups have attempted to address this multi-path system
theoretically, and predicted the Fano resonance in the
parallel-coupled DQD
system.\cite{mahan,kang,guevara,baizhiming,bingdong,guevara2}
However, it seems that a physically transparent picture for the
Fano effect in the DQD system is still lacking. For example, it is
unclear what the resonant and nonresonant channels are in this
Fano system. Moreover, a systematic study is required for
exploring various possibilities of tuning Fano effect with
external fields. In this paper, we intend to provide with a
natural yet simple explanation for the Fano effect in the
parallel-coupled DQD, and propose several ways to control the Fano
resonance in the conductance spectra by the electrostatic and
magnetic approaches.

 The paper is organized as follows. In Sec.\ref{sec:model},
a widely used two-level Fano-Anderson model is introduced with an
inter-dot coupling term added. Since the coupled quantum dots may
be considered as an artificial QD-molecules,\cite{livermore} an
effective Hamiltonian in terms of the bonding and antibonding
states of the QD-molecule may form an appropriate working basis.
Thus with the help of the Green function technique and the
equation of motion method,\cite{haug} the density of states is
calculated in three asymmetric configurations classified according
to the spatial symmetry of the dot-lead coupling.\cite{konig} In
Sec. \ref{sec:calculation}, the conductance formula is derived for
this system,\cite{datta,meir} whereby the Fano lineshape in
conductance spectra is calculated in the absence of the magnetic
flux. In Sec.\ref{sec:explain}, a simple mechanism explaining the
Fano lineshape in the DQD conductance spectrum is presented. Then,
the possibilities of tuning the conductance lineshape by various
electrostatic and magnetic methods are described in details, and
several novel effects are predicted. Most importantly, by tuning
the the total magnetic flux, or the flux difference between the
left and right parts of the AB ring, the swap effect between two
resonance peaks in the conductance spectra is predicted, which
might be of potential application as a type of C-Not gate in the
quantum computation. Finally, a brief summary is drawn and
presented.

\section{\label{sec:model}Physical Model}

\begin{figure}[htbp]
\centering
\includegraphics[width=0.45\textwidth]{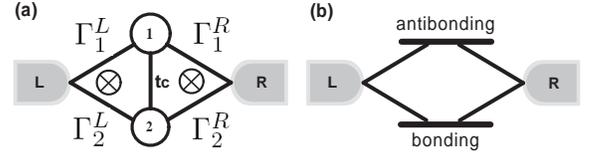}
\caption{(a) Schematic diagram for a tunneling-coupled parallel
DQD system. (b) Schematic diagram of the parallel-coupled DQD
system in the molecular orbital representation.}\label{fig:model}
\end{figure}

We start with the Fano-Anderson model for the parallel-coupled DQD
where the discrete states in two quantum dots are coupled each
other via tunneling. (Fig.\ref{fig:model}a) Then the Hamiltonian
reads
\begin{equation}\label{hamiltonian}
H=H_{leads}+H_{DD}+H_T.
\end{equation}
The $H_{leads}$ in Eq.(\ref{hamiltonian}) represents the
noninteracting electron gas in the left(L) and right(R) leads,
\begin{equation}
H_{leads}=\sum_{k,\alpha=L,R}\varepsilon_{k\alpha}c_{k\alpha}^{\dagger}c_{k\alpha},
\end{equation}
where, $c_{k\alpha}^{\dagger}, c_{k\alpha}$ are the creation and
annihilation operators for a continuum in the lead $\alpha$ with
energy $\varepsilon_{k\alpha}$. The $H_{DD}$ in
Eq.(\ref{hamiltonian}) describes the QD electrons and their mutual
coupling in the DQD, {\it i.e.}
\begin{equation}\label{hd1}
H_{DD}=\sum_{i=1,2}\varepsilon_i d_i^{\dagger}d_i -
t_ce^{i\theta}d_1^{\dagger}d_2 - t_ce^{-i\theta}d_2^{\dagger}d_1.
\end{equation}
The first term in Eq.(\ref{hd1}), $d_i^{\dagger}$ $(d_i)$,
represents the create (annihilate) operator of the electron with
the energy $\varepsilon_i$ in the dot $i$. The second and third
terms in Eq.(\ref{hd1}) denote the inter-dot coupling, in which
$t_c$ is the coupling strength taken as a real parameter, and
$\theta$ denotes a phase shift related to the flux difference
between the left and right sub-rings.
The $H_{T}$ in Eq.(\ref{hamiltonian}) represents the tunneling
coupling between the QD and lead electrons,
\begin{eqnarray}
H_T=&&\sum_{k,{\alpha}=L,R}\sum_{i=1,2} V_{\alpha
i}d_i^{\dagger}c_{k{\alpha}} +  h.c.,
\end{eqnarray}
where the tunneling matrix element
$V_{L1}=|V_{L1}|e^{i\frac{\phi}{4}},
V_{L2}^*=|V_{L2}|e^{i\frac{\phi}{4}},
V_{R1}^*=|V_{R1}|e^{i\frac{\phi}{4}}$, and
$V_{R2}=|V_{R2}|e^{i\frac{\phi}{4}}$. Here for the sake of
simplicity, $V_{\alpha i}$ is assumed to be independent of $k$,
and the phase shift due to the total magnetic flux threading into
the AB ring, $\phi$, is assumed to distribute evenly among 4
sections of the DQD-AB ring. Namely, $\phi=
2\pi(\Phi_R+\Phi_L)/\Phi_0$, where the flux quantum $\Phi_0=hc/e$.
Thus, $\theta= \pi(\Phi_R-\Phi_L)/\Phi_0$. In the following
calculation, we define the line-width matrix as
$\Gamma^{\alpha}_{ij}=\sum_{k}V_{\alpha i}V^*_{\alpha j}2\pi
\delta(\varepsilon-\varepsilon_{k{\alpha}})$ (${\alpha}=L,R)$ and
$\mathbf{\Gamma}=\mathbf{\Gamma}^L+\mathbf{\Gamma}^R$. According
to Fig.\ref{fig:model}(a), the line-width matrices in the QD
representation read
\begin{equation}\label{gstrgammaL}
\mathbf{\Gamma}^L=\left(\begin{array}{cc}
 \Gamma^L_{1} &
\sqrt{\Gamma^L_{1}\Gamma^L_{2}}e^{i\frac{\phi}{2}}\\
\sqrt{\Gamma^L_{1}\Gamma^L_{2}}e^{-i\frac{\phi}{2}} &\Gamma^L_{2}
\end{array}\right),\nonumber
\end{equation}
and
\begin{equation}\label{gstrgammaR}
\mathbf{\Gamma}^R=\left(\begin{array}{cc}
 \Gamma^R_{1} &
\sqrt{\Gamma^R_{1}\Gamma^R_{2}}e^{-i\frac{\phi}{2}}\\
\sqrt{\Gamma^R_{1}\Gamma^R_{2}}e^{i\frac{\phi}{2}} &\Gamma^R_{2}
\end{array}\right),
\end{equation}
where $\Gamma^{\alpha}_i$ is short for $\Gamma^{\alpha}_{ii}$ .

To make the physical picture clearer and formalism simpler, it is
attractive to introduce a QD-molecule representation by
transforming two tunneling-coupled QD levels into the bonding and
antibonding states of the QD-molecule. The operator for a molecule
state can be expressed as a linear superposition of the QD
operators as
\begin{equation}\label{unitaray}
\left(\begin{array}{c}
f_+\\f_-\end{array}\right)=\left(\begin{array}{cc}
 \cos\beta e^{-i\theta} &
-\sin\beta\\
\sin\beta &\cos\beta e^{i\theta}
\end{array}\right)\left( \begin{array}{c}d_1\\d_2\end{array}\right),
\end{equation}
where, $f_-$ and $f_+$ are referred to as the annihilation
operators for the bonding and antibonding states of the artificial
QD-molecule. And the parameter $\beta$ is defined as
${\beta}=1/2\tan^{-1}[2t_c/(\varepsilon_1-\varepsilon_2)]. $ For
mathematical simplicity, in the following only the symmetric case
is studied, {\it i.e.}
$\varepsilon_1=\varepsilon_2=\varepsilon_0$, and thus
$\beta=\pi/4$. Then the Hamiltonian for coupled dots is decoupled
as
\begin{equation}\label{Hd2}
\widetilde{H}_{DD}=(\varepsilon_0+t_c)f_+^{\dagger}f_+
+(\varepsilon_0-t_c)f_-^{\dagger}f_-,
\end{equation}
In the molecular-state representation, the tunneling Hamiltonian
between the leads and DQD is rewritten into
\begin{eqnarray}\label{Ht2}
\widetilde{H}_T=\sum_{k,{\alpha}=L,R}\sum_{i=+,-}\widetilde{V}_{\alpha
i}f^{\dagger}_i c_{k{\alpha}}+h.c.,
\end{eqnarray}
where the effective tunneling matrix elements are
\begin{equation}\label{efftun}
\left(\begin{array}{c} \widetilde{V}_{\alpha
+}\\\widetilde{V}_{\alpha
-}\end{array}\right)=\frac{1}{\sqrt{2}}\left(\begin{array}{cc}
 e^{-i\theta} &
-1\\
1 & e^{i\theta}
\end{array}\right)\left( \begin{array}{c}V_{\alpha
1}\\V_{\alpha 2}\end{array}\right).
\end{equation}
Now, the DQD system has been mapped into a system of two
independent molecular states, which are connected to leads,
respectively({\it cf.} Fig.\ref{fig:model}b). In the
molecular-state representation, the linewidth matrices read
\begin{equation}\label{mogamma}
\widetilde{{\mathbf\Gamma}}_{ij}^{\alpha}=\sum_{k}\widetilde{V}_{\alpha
i}\widetilde{V}^*_{\alpha j} 2\pi {\delta} (
{\varepsilon}-\varepsilon_{k{\alpha}} ),
\end{equation}
i.e.
\begin{widetext}
\begin{equation}\label{mogammaL}
\widetilde{\mathbf{\Gamma}}^L=\frac{1}{2}\left(\begin{array}{cc}
 \Gamma^L_{1}+\Gamma^L_{2}-2\sqrt{\Gamma^L_{1}\Gamma^L_{2}}\cos(\frac{\phi}{2}-\theta) &
(\Gamma^L_{1}-\Gamma^L_{2})e^{-i\theta}+\sqrt{\Gamma^L_{1}\Gamma^L_{2}}e^{i(\frac{\phi}{2}-2\theta)}-\sqrt{\Gamma^L_{1}\Gamma^L_{2}}e^{-i\frac{\phi}{2}}\\
(\Gamma^L_{1}-\Gamma^L_{2})e^{i\theta}+\sqrt{\Gamma^L_{1}\Gamma^L_{2}}e^{-i(\frac{\phi}{2}-2\theta)}-\sqrt{\Gamma^L_{1}\Gamma^L_{2}}e^{i\frac{\phi}{2}}
&\Gamma^L_{1}+\Gamma^L_{2}+2\sqrt{\Gamma^L_{1}\Gamma^L_{2}}\cos(\frac{\phi}{2}-\theta)
\end{array}\right),
\end{equation}
and
\begin{equation}\label{mogammaR}
\widetilde{\mathbf{\Gamma}}^R=\frac{1}{2}\left(\begin{array}{cc}
 \Gamma^R_{1}+\Gamma^R_{2}-2\sqrt{\Gamma^R_{1}\Gamma^R_{2}}\cos(\frac{\phi}{2}+\theta) &
(\Gamma^R_{1}-\Gamma^R_{2})e^{-i\theta}+\sqrt{\Gamma^R_{1}\Gamma^R_{2}}e^{-i(\frac{\phi}{2}+2\theta)}-\sqrt{\Gamma^R_{1}\Gamma^R_{2}}e^{i\frac{\phi}{2}}\\
(\Gamma^R_{1}-\Gamma^R_{2})e^{i\theta}+\sqrt{\Gamma^R_{1}\Gamma^R_{2}}e^{i(\frac{\phi}{2}+2\theta)}-\sqrt{\Gamma^R_{1}\Gamma^R_{2}}e^{-i\frac{\phi}{2}}
 &\Gamma^R_{1}+\Gamma^R_{2}+2\sqrt{\Gamma^R_{1}\Gamma^R_{2}}\cos(\frac{\phi}{2}+\theta)
\end{array}\right).
\end{equation}
\end{widetext}

To estimate the broadening of the molecular level due to its
couplings to leads, let us first calculate the density of
states(DOS) for each state. The retarded Green function for the
molecular state is defined as $ G^r_{\pm}=-i\theta(t)\langle
\{f_{\pm}(t),f^{\dagger}_{\pm}\}\rangle $. With the equation of
motion approach\cite{haug}, we have
\begin{equation}\label{gfmo3}
G^r_{\pm}(\varepsilon)=\frac{1}{\varepsilon-(\varepsilon_0\pm
t_c)+i(\Gamma_{\pm})},
\end{equation}
where the imaginary part of the self-energy is equal to
\begin{eqnarray}\label{gfmo5}
\Gamma_{\pm}=\frac{1}{2}(\widetilde{\Gamma}^L_{\pm\pm}+\widetilde{\Gamma}^R_{\pm\pm})
\end{eqnarray}
The local density of states is defined as the imaginary part of
the retarded Green function as
\begin{equation}\label{rhopm}
\rho_{\pm}(\varepsilon)=-(1/\pi)\mathrm{Im}G^r_{\pm}(\varepsilon)
=\frac{1}{\pi}\frac{\Gamma_{\pm}}{[\varepsilon-(\varepsilon_0 \pm
t_c)]^2+(\Gamma_{\pm})^2},
\end{equation}
which is the Lorentzian peaked at the molecular level.

\begin{figure}[htbp] \centering
\includegraphics[width=0.45\textwidth]{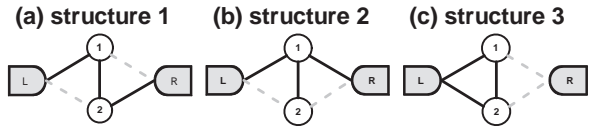}
\caption{Three structures investigated, in which the solid and
dash lines stand for the stronger and weaker tunnel coupling,
respectively.}\label{fig:structure}
\end{figure}

Three configurations for the DQD system according to the
asymmetric couplings between two dots and leads\cite{konig} that
we focus on are shown in Fig.\ref{fig:structure}: (a)structure 1,
in which
$\Gamma^L_{1}=\Gamma^R_{2}=2\Gamma^L_{2}=2\Gamma^R_{1}=\gamma$;
(b) structure 2, in which
$\Gamma^L_{1}=\Gamma^R_{1}=2\Gamma^L_{2}=2\Gamma^R_{2}=\gamma$;
and (c) structure 3, where
$\Gamma^L_{1}=\Gamma^L_{2}=2\Gamma^R_{1}=2\Gamma^R_{2}=\gamma$.
Here the $\gamma$ is taken as an energy unit. These typical
structures for the DQD system are basic, yet convenient to analyze
theoretically.

\begin{figure}[htbp]
\centering
\includegraphics[width=0.5\textwidth]{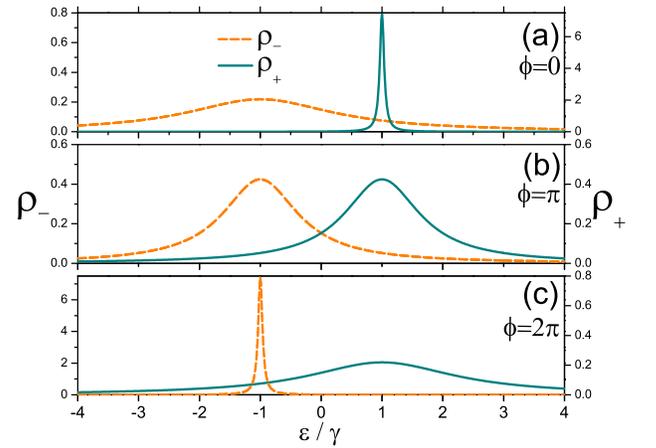}
\caption{(Color online)The density of states for two molecular
states in structure 1 and structure 2 as shown in
Fig.\ref{fig:structure}, where $t_c=\gamma$ which is taken as an
energy scale.}\label{fig:dos}
\end{figure}

Fig.\ref{fig:dos} shows how the calculated DOS of two molecular
states changes with the total magnetic flux $\phi$ in the
structure 1 and 2. It is noticed that the structure 1 and 2 share
the identical DOS. Two points are worth to pointing out. Firstly,
since the full width at half maximum $2\Gamma_{\pm}$ for each
molecular state depends on the parameters
 $\Gamma^{\alpha}_{i}$, $\phi$, and $\theta$, the broadenings of
molecular states could be tuned not only by the dot-lead coupling
strength and total magnetic flux, but also by the flux difference.
Secondly, the broadening of one molecular state is always
accompanied by the shrinking of the other molecular state because
the trace of the matrix of $(\widetilde{\mathbf{\Gamma}}^L+
\widetilde{\mathbf{\Gamma}}^R)$ is an invariant, as explicitly
shown by Eqs. (\ref{mogammaL}) and (\ref{mogammaR}). Thirdly, in
the absence of the magnetic flux ($\phi=0$),
$\Gamma_{+}\rightarrow 0^+$ in the structure 3, i.e. the
antibonding state is totally decoupled from the leads and
possesses an infinite lifetime. The finite $\phi$ introduced by
the magnetic flux in this structure results in the finite coupling
between the antibonding state and leads, and thus a finite width
for the antibonding state.

\section{\label{sec:calculation}Conductance}
To study the transport through the DQD system, the conductance at
zero temperature is derived, which, on the basis of the
non-interacting molecular levels, can be reduced to the
Landauer-B\"{u}ttiker formula \cite{datta}
\begin{equation}\label{conductance}
\mathcal{G}(\varepsilon)=\frac{2e^2}{h}T(\varepsilon),
\end{equation}
where, $\varepsilon$ is the Fermi energy at both leads at
equilibrium. In the absence of Coulomb interaction between the
electrons on the dots, the transmission $T(\varepsilon)$ is
expressed as\cite{meir}
\begin{equation}\label{transmission}
T(\varepsilon)=\mathrm{Tr}[\mathbf{G}^a(\varepsilon)\mathbf{\Gamma}^R
\mathbf{G}^r (\varepsilon)\mathbf{\Gamma}^L],
\end{equation}
where the retarded and advanced Green functions, $\mathbf{G}^r$
and $\mathbf{G}^a$, and the linewidth matrices
$\mathbf{\Gamma}^{L(R)}$ can be expressed in molecular states
representation or in dot levels representation, respectively.

The retarded Green function in the dot level representation is
defined as $G^r_{ij}=-i\theta(t)\langle \{
d_i(t),d^{\dagger}_j\}\rangle$, and its Fourier transformation
$G^r_{ij}(\varepsilon)\equiv\langle\langle
d_i|d^{\dagger}_j\rangle\rangle$ satisfies
\begin{equation}\label{eom}
\langle\langle d_i|d^{\dagger}_j\rangle\rangle=\delta_{ij}
+\langle\langle[d_i,H]|d^{\dagger}_j\rangle\rangle,
\end{equation}
which generates a closed set of linear equations for
$G^r_{ij}(\varepsilon)$. The solution of $\mathbf{G}^r$ is given
by
\begin{eqnarray}\label{Grdot}
&&\mathbf{G}^r(\varepsilon)=\nonumber\\
&&\left(%
\begin{array}{cc}
  \varepsilon-\varepsilon_1+\frac{i}{2}(\Gamma^L_{11}+\Gamma^R_{11}) &
  t_c+\frac{i}{2}(\Gamma^L_{12}+\Gamma^R_{12}) \\
  t_c^*+\frac{i}{2}(\Gamma^L_{21}+\Gamma^R_{21}) &
  \varepsilon-\varepsilon_2+\frac{i}{2}(\Gamma^L_{22}+\Gamma^R_{22}) \\
\end{array}%
\right)^{-1}.\nonumber\\
\end{eqnarray}
The advanced Green function is the Hermite conjugate of the
retarded Green function. Put Eq.(\ref{Grdot}) into
Eq.(\ref{transmission}) and after some algebra, the transmission
probability through the DQD system is found to be
\begin{eqnarray}\label{conduct1}
&&T(\varepsilon)=\frac{a(\varepsilon-\varepsilon_0)^2
+b(\varepsilon-\varepsilon_0)+c}{\{[\varepsilon-(\varepsilon_0+t_c)]^2+\Gamma^2_+\}
\{[\varepsilon-(\varepsilon_0-t_c)]^2+\Gamma^2_-\}},\nonumber\\
\end{eqnarray}
where
\begin{eqnarray}\label{abc}
a&=&\Gamma^L_1\Gamma^R_1+\Gamma^L_2\Gamma^R_2
+2\cos\phi\sqrt{\Gamma^L_1\Gamma^R_1\Gamma^L_2\Gamma^R_2},\nonumber\\
b&=&-2t_c(\Gamma^L_1+\Gamma^L_2)\sqrt{\Gamma^R_1\Gamma^R_2}\cos(\theta+\frac{\phi}{2})\nonumber\\
&&-2t_c(\Gamma^R_1+\Gamma^R_2)\sqrt{\Gamma^L_1\Gamma^L_2}\cos(\theta-\frac{\phi}{2}),\nonumber\\
c&=&t_c^2(\Gamma^L_1\Gamma^R_2+\Gamma^L_2\Gamma^R_1
+2\cos 2\theta\sqrt{\Gamma^L_1\Gamma^L_2\Gamma^R_1\Gamma^R_2}\ ).\nonumber\\
\end{eqnarray}
 The conductance through the DQD system as expressed in Eq.(\ref{conduct1}) depends in
general on the energy levels of two dots, dot-lead coupling,
 inter-dot coupling, and phase shift induced by the magnetic flux.
Without the magnetic flux, i.e. $\phi=0$, and $\theta=0$,
Eq.(\ref{conduct1}) recovers to the result by Guevara \emph{et
al}.\cite{guevara}; while in the limit of vanishing inter-dot
coupling ($t_c=0$), the result of Kubala \emph{et
al}.\cite{kubala} is repeated.

In the molecular state representation, the total conductance can
also be divided into
\begin{equation}\label{totalconduct}
\mathcal{G}_{\mathrm{total}}(\varepsilon)=\frac{2e^2}{h}\{T_{+}+T_{-}+T_{\mathrm{inter}}\},
\end{equation}
where the transmission via each molecular state is
\begin{eqnarray}\label{Tmole}
T_{\pm}(\varepsilon)=\frac{\widetilde{\Gamma}^L_{\pm\pm}\widetilde{\Gamma}^R_{\pm\pm}}
{[\varepsilon-(\varepsilon_0 \pm t_c)]^2+\Gamma_{\pm}^2},
\end{eqnarray}
and the interference term is given by
\begin{eqnarray}\label{gcoherent}
T_{\mathrm{inter}}=2\mathrm{Re}\{\frac
{\widetilde{\Gamma}^R_{+-}\widetilde{\Gamma}^L_{-+}}
{[\varepsilon-(\varepsilon_0 +
t_c)-i\Gamma_{+}][\varepsilon-(\varepsilon_0- t_c)+i\Gamma_{-}]}
\}.\nonumber\\
\end{eqnarray}



\section{\label{sec:explain}Tunable Fano effect}

\begin{figure}[htbp]
\centering
\includegraphics[width=0.5\textwidth]{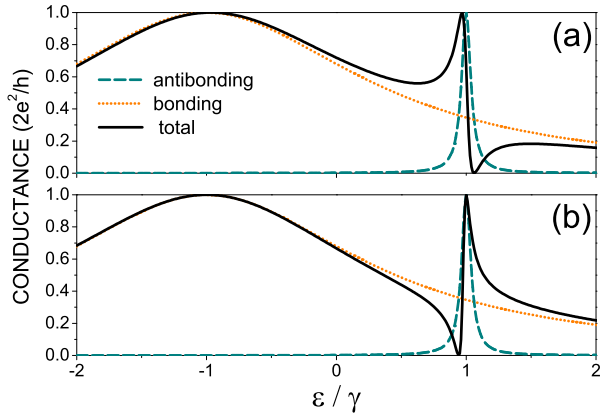}
\caption{(Color online)Conductance spectra in structure 1(a) and
structure 2(b). The parameters for calculations are $\theta=0,
\phi=0, \varepsilon_0=0$, and $t_c=\gamma$. Notice that the
conductance spectra in two structures have quite different
interference pattern though the identical DOS as in
Fig.\ref{fig:dos}. }\label{fig:fano3}
\end{figure}

As shown in Fig.\ref{fig:fano3}, the total conductance through the
parallel-couple DQD system consists of a Breit-Wigner peak and a
Fano peak, which is quite different from the DOS where two
Lorentzians are superposed. Based on the molecular level
representation formulated above, let us first explain how the Fano
interference is produced in somewhat details, then show how to
tune it. Although similar observations have been reported in
Refs.\onlinecite{kang,guevara,guevara2,baizhiming}, it seems that
till now no simple and transparent explanation is available on the
formation of the Fano lineshape in this structure.

When a discrete level is buried into a continuum, the coupling
between the discrete and continuous states gives rise to the
renormalization of the states of whole system. The phase of the
renormalized wave function varies by $\sim \pi$ swiftly as the
energy transverses an interval $\sim \Gamma$ around the discrete
level, where $\Gamma$ is the broadening of the discrete level due
to coupling with the continua.\cite{fano,mahan} Then, if there is
a reference channel whose phase changes little in the interval of
$\Gamma$ around the discrete level, the quantum interference above
and below the resonance level will be responsible for the Fano
lineshape in the conductance spectra, for example, the
experimentally observed Fano lineshape in the hybrid system of a
QD and a reference arm.\cite{Kobayashi2002,Kobayashi2003,johnson}

In the present DQD multi-path system, it is not simple to identify
the resonant and the reference channel. But, on the basis of the
molecular level representation, it is straightforward to interpret
the Fano resonance in the DQD system in terms of the interference
between two channels.

Usually, two molecular levels are coupled to the leads unequally.
In the absence of magnetic flux, Eq.(\ref{gfmo5}) simply reduces
to
\begin{equation}\label{Gamma_pm}
\Gamma_{\pm}=\frac{1}{4}(\Gamma^L_{1}+\Gamma^L_{2}+\Gamma^R_{1}+\Gamma^R_{2})
\mp
\frac{1}{2}(\sqrt{\Gamma^L_{1}\Gamma^L_{2}}+\sqrt{\Gamma^R_{1}\Gamma^R_{2}}),
\end{equation}
i.e., the broadening of one level is always accompanied by the
shrinking of the other. The molecular level associated with a
wider band can be referred to as the strongly-coupled one, while
that with narrow band is referred to as the weakly-coupled level.
Suppose that the broadening of the strongly-coupled level entirely
covers the band width with the weakly-coupled level, and the phase
shift for the strongly-coupled channel is negligibly small around
the weakly-coupled level, then a phase shift of $\pi$ across the
weakly-coupled level can be detected with characteristic of the
Fano lineshape. Namely, the waves through two channels interfere
constructively for electron with energy on one side of the
weakly-coupled level; while interfere destructively on the other
side. As a result, the Fano lineshape shows up around the
weakly-coupled state. It is worth pointing out that the $\pi$
phase shift also happens around the strongly-coupled level, but on
a energy scale much larger than that around the weakly-coupled
level (Fig.\ref{fig:dos}). That is why usually there is only one
Fano peak around the weakly-coupled level, and the other peak
around the strongly-coupled level only exhibits little asymmetry.

Compared with the Fano effect in the hybrid
system,\cite{Kobayashi2002,Kobayashi2003,johnson} where the
nonresonant channel served by a quantum point contact detects the
$\pi$ phase shift around the resonant tunneling channel through
the QD, in the DQD system, the reference channel is the "less
resonant" tunneling channel on one side of the strongly-coupled
level, and the other "more resonant" channel through the
weakly-coupled level is accompanied with a swift $\pi$ phase shift
within a small energy region.

The Fano lineshape in the present system can be tuned by the
applied electrostatic and magnetic fields. Compared with the one
dot and one arm case, where the magnetic field affects only the
phase of the electron, but not the interaction
strength,\cite{Kobayashi2002,Kobayashi2003} in the
parallel-coupled DQD, not only the phase but also the magnitude of
the effective coupling between the molecular states and leads can
be tuned by the magnetic flux. Of course, one should keep in mind
that the inter-dot coupling is the prerequisite to the tunable
Fano effect.
\begin{figure}[htbp]
\centering
\includegraphics[width=0.5\textwidth]{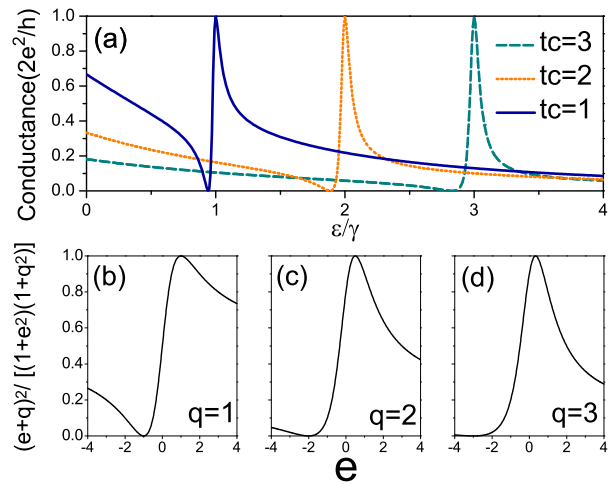}
\caption{(Color online)(a)Calculated conductance in the structure
2 for three different values of inter-dot coupling $t_c$, where
the Fermi energy is roughly at the antibonding level. (b),(c),and
(d) Simulation of the normalized Fano lineshape with different
asymmetric factors $q$ as compared to the results obtained by
Green functions in (a).}\label{fig:tuningtc}
\end{figure}
\subsection{Electrostatic tuning}
The inter-dot coupling $t_c$ can be tuned by adjusting the height
and thickness of the tunneling barrier between two dots through
gate voltages. So can be done the dot-lead coupling
$\Gamma^{L(R)}_{i}$.

The interference between the resonance and the reference channels
can be described as
\begin{equation}\label{Fanolineshape}
|t_R\frac{\Gamma}{\omega-\epsilon_0+i\Gamma}+t_N
e^{i\phi}|^2=t^2_N\frac{|\widetilde{\epsilon}+q|^2}{\widetilde{\epsilon}^2+1},
\end{equation}
where $t_R$ and $t_N$ denote respectively the transmission
amplitude via the resonance and the reference channels, the
detuning $\widetilde{\epsilon}=(\omega-\epsilon_0)/\Gamma$, and
the asymmetric factor $q=i+ t_R e^{-i\phi}/t_N $. It is well known
that when the asymmetric factor $q$ is small,
Eq.(\ref{Fanolineshape}) gives rise to the asymmetric Fano
lineshape; while for large $q$, it tends to the symmetric
Lorentzian. Fig. \ref{fig:tuningtc}(a) depicts the conductance
spectra for three different inter-dot coupling strengths in
Structure 2. As for comparison, the normalized Fano lineshape
(multiplied by a normalized factor $1/(1+q^2)$) is also plotted in
Fig.\ref{fig:tuningtc}(b),(c), and (d), demonstrating how the Fano
lineshape evolves with increasing $q$. It is obvious that as the
inter-dot coupling increases, the bonding and antibonding
splitting increases. As a result, the amplitude through the
channel associated with the strongly-coupled level but at the
energy range of antibonding peak, $t_N$, decreases, and since $q$
is inversely proportional to $t_N$, the asymmetric factor $q$
increases. This further confirms our explanation above about the
origin of Fano effect in this parallel-coupled DQD system.

\begin{figure}[htbp]
\centering
\includegraphics[width=0.4\textwidth]{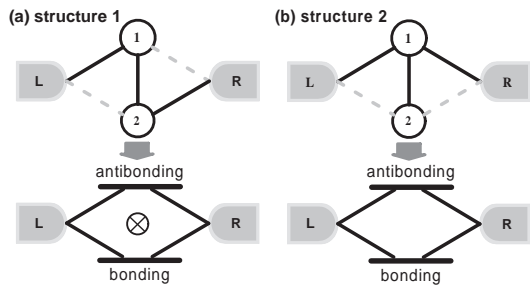}
\caption{(a) Structure 1 and (b)structure 2 can be cast into two
different models of molecular states. The flux through the loop of
structure 1 differs from that of structure 2 by a phase of
$\pi$.}\label{fig:jump}
\end{figure}

When tuning the dot-lead coupling strength $\Gamma^{L(R)}_{i}$ by
tuning the gate voltage, the structure studied is changed, and the
Fano lineshape is changed accordingly. For example, when the
strong and weak coupling in the linewidth matrix is adjusted such
that the structure 1 is transformed into the structure 2, (see
Fig.\ref{fig:fano3}b), noticeably, the tail direction of the Fano
peak is flipped. This delicate change can be understood in terms
of the product of the effective tunneling matrix elements
\begin{eqnarray}\label{flip}
&&\widetilde{V}_{L+}\widetilde{V}^*_{R+}\widetilde{V}_{R-}\widetilde{V}^*_{L-}\nonumber\\
=&&\frac{1}{4}(V_{L1}-V_{L2})(V_{R1}-V_{R2})(V_{L1}+V_{L2})(V_{R1}+V_{R2}).\nonumber\\
\end{eqnarray}
In the wide band limit, a linewidth matrix element is proportional
to the product of two dot-lead tunneling matrix elements.
According to Fig.\ref{fig:structure}, the expression above for the
structure 1 differs from that for structure 2 by a minus sign,
which implies that, compared to structure 2, an extra flux of
$\pi$ threads the loop in the structure 1(Fig.\ref{fig:jump}).
Thus, the Fano lineshape in structure 1 is just opposite to that
in structure 2, i.e., if two channels interfere with each other
constructively in structure 1, then destructively in structure 2;
and vice versa.
\subsection{\label{sec:mag}Magnetic flux tuning and swap operation}
Coupled quantum dot systems have been proposed to materialize
quantum bit for quantum computation\cite{loss}. The swap operation
is an important element to the controlled-NOT gate, which is the
key to implementing the quantum computation. \cite{divincenzo}
Recently, a new mechanism has been proposed to realize the swap
operation in the parallel-coupled DQD system by using the
time-dependent inter-dot spin superexchange J(t), which flips the
singlet and triplet states formed by two localize
electrons.\cite{zhang} In the following we will discuss a new type
of swap operation in the DQD system, which flips two quantum
states by tuning the magnetic flux, including tuning the total
flux $\phi$ or the flux difference $2\theta$ between the left and
right sub-rings. For the purpose of comparison, only tuning one
parameter, and letting the other alone.
\begin{figure}[htbp] \centering
\includegraphics[width=0.5\textwidth]{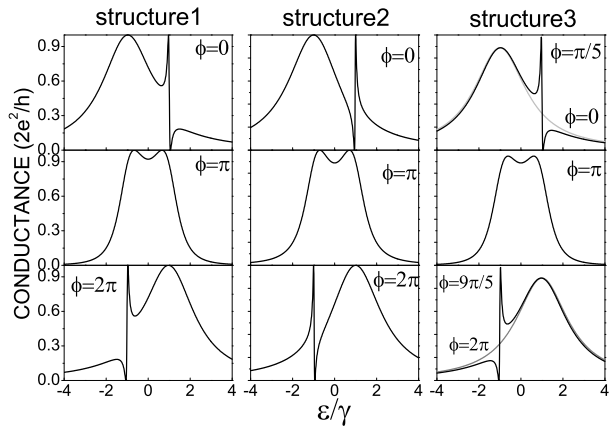}
\caption{The evolution of transmission spectrum with total
magnetic flux $\phi$ tuned for three asymmetric configurations.
}\label{fig:tuningphi}
\end{figure}

First, we tune the total flux $\phi$ and let $\theta=0$. When
$\phi=(2n+1)2\pi$ (n is an integer), Eq.(\ref{gfmo5}) turns out to
be
$\Gamma_{\pm}=\frac{1}{4}(\Gamma^L_{1}+\Gamma^L_{2}+\Gamma^R_{1}+\Gamma^R_{2})
\pm
\frac{1}{2}(\sqrt{\Gamma^L_{1}\Gamma^L_{2}}+\sqrt{\Gamma^R_{1}\Gamma^R_{2}}).$
Compared with the case of null flux, or $\phi=4n\pi$[
Eq.(\ref{Gamma_pm})], the widths of the bonding and antibonding
states are interchanged. Fig.\ref{fig:tuningphi} demonstrates the
evolution of conductances with changing the total magnetic flux in
three asymmetric configurations for zero $\theta$. In this
circumstance the Fano and  Breit-Wigner peaks in the conductance
spectra of the structures 1 and 2 have been exchanged each other,
when changing $\phi$ from $(2n+1)2\pi$ to $\phi=4n\pi$. On the
other hand, when $\phi=(2n+1)\pi$, two identical peaks appear in
conductance spectra symmetrically, which is quite similar to the
DOS spectra. In structure 3, no Fano peak exists without the
magnetic flux according to Eq.(\ref{rhopm}) and
Fig.\ref{fig:tuningphi}, because the antibonding state is totally
decoupled from the leads. However, if $\phi$ deviates a little
from zero, or more generally, from a multiple of $2\pi$, the
channel connecting the decoupled state and the leads is open, and
the Fano interference comes back again. Our model calculation
indicates that, this flux-dependent Fano effect takes place
whenever $\Gamma^L_{1}=\Gamma^L_{2}$ and
$\Gamma^R_{1}=\Gamma^R_{2}$, regardless whether four dot-lead
couplings are the identical.\cite{guevara}

\begin{figure}[htbp]
\centering
\includegraphics[width=0.5\textwidth]{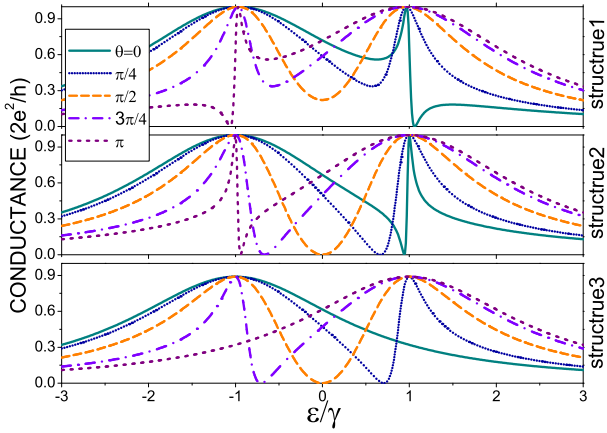}
\caption{(Color online)The evolution of transmission spectrum with
phase difference $\theta$ in three asymmetric configurations.
}\label{fig:tuningtheta}
\end{figure}

Tuning $\theta$, the half of the flux difference between the left
and right sub-rings, can also lead to two resonance peak swap and
the flux-dependent Fano effect as shown in
Fig.\ref{fig:tuningtheta}. It is noticed when $\theta=\pi$, the
results coincide with what obtained when $\phi=2\pi$.

\begin{figure}[htbp] \centering
\includegraphics[width=0.4\textwidth]{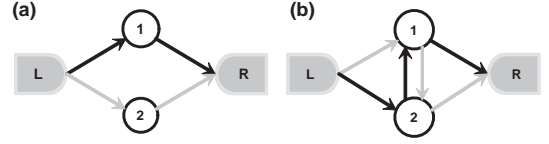}
\caption{Illustration for two different periods of conductance.
Multi-pathway interference as functions of (a) the total flux
$\phi$, and (b) the flux difference
$\theta$.}\label{fig:multipathway}
\end{figure}

It is interesting to note that the conductance oscillation is a
periodic function of $\phi$ with a period of $4\pi$, and a
periodic function of $\theta$ with the period of $2\pi$. The
difference in the period can be readily explained as the
multi-pathway nature in this parallel-coupled DQD system. If an
electron transits from the left to the right, it gathers a phase
factor $e^{i(\frac{\phi}{4}+\frac{\phi}{4})}=e^{i\frac{\phi}{2}}$
via the upper arms and $e^{-i\frac{\phi}{2}}$ via the lower
arms(Fig.\ref{fig:multipathway}a). The interference of the two
paths gives $\cos\frac{\phi}{2}$, which is associated with a
period of $4\pi$. On the other hand, if two routes of an electron
are illustrated as the arrows in Fig.\ref{fig:multipathway}b, the
interference yields $\cos{\theta}$ and the $2\pi$ periodicity,
because the first path from the left lead $\rightarrow$ dot 1
$\rightarrow$ dot 2 $\rightarrow$ right lead accumulates a phase
$e^{i(\frac{\phi}{4} + \theta - \frac{\phi}{4})}=e^{i\theta}$, and
the second symmetric route through dot2 $\rightarrow$ dot1 gives a
phase $e^{-i\theta}$.

\begin{figure}[htbp] \centering
\includegraphics[width=0.5\textwidth]{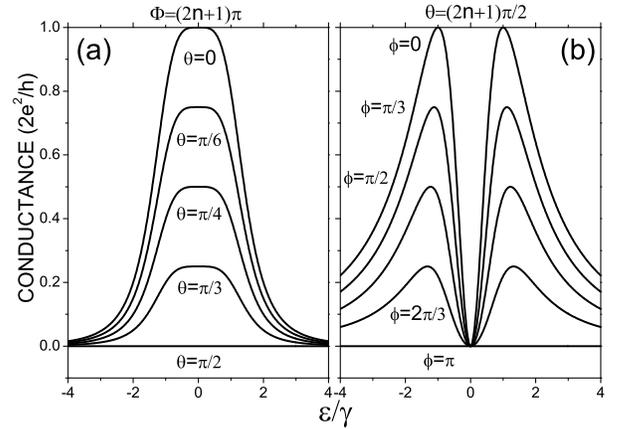}
\caption{Continuously modulated conductance by the magnetic flux.
(a) The conductance is tuned by $\theta$, when $\phi$ is fixed at
$(2n+1)\pi$; (b) The conductance is tuned by $\phi$, when
$\theta=(2n+1)\frac{\pi}{2}$.}\label{fig:switch}
\end{figure}

The quantum interference tuned by the magnetic flux in the DQD
system makes the conductance to vary in some fancy way. Let us
look at a peculiar instance, the continuous manipulation of the
conductance with the lineshape retained. In Fig.\ref{fig:switch},
two cases of the flux tuning are: (1) $\phi=(2n+1)\pi$,
$\theta\in[0,\frac{\pi}{2}]$; and (2)
$\theta=(2n+1)\frac{\pi}{2}$, $\phi\in[0,\pi]$. According to
Eq.(\ref{mogammaL}) and Eq.(\ref{mogammaR}), once
$\theta=(2n+1)\frac{\pi}{2}$ or $\phi=(2n+1)\pi$, the lineshape of
the DOS for two molecular levels is fixed. In our calculations,
$\Gamma^L_1=\Gamma^L_2=\Gamma^R_1=\Gamma^R_2=t_c=\gamma,\
\varepsilon_0=0$, only a single peak appears in
Fig.\ref{fig:switch}(a)(when increasing $t_c$, the conductance
will recover the double-peak feature). In contrast, in
Fig.\ref{fig:switch}(b) due to the destructive interference, the
conductance at $\varepsilon_0$ is always zero. Besides, a fully
symmetric configuration with four identical dot-lead couplings is
considered in this calculation, so that the maximum and minimum
values of the conductance are precisely at $2e^2/h$ and 0,
respectively.

\section{Conclusions}
In summary, the transport through the parallel-coupled DQD system
has been studied, in which a particular attention is paid to the
mechanism of the Fano lineshape in conductance spectra as well as
its tunability. Due to the inter-dot coupling, a QD-molecule is
formed and can be the proper representation for the present
investigations. Due to the coupling between the molecular levels
and leads, two levels are broadened into two bands: one is wider
which is associated with the strongly-coupled level, and the other
is narrower band related to the weakly-coupled level. When the
wider band covers the narrower band, and the phase shift of the
wavefunction of the wider band is negligible around the
weakly-coupled level, the $\pi$ phase shift at resonance of the
wavefunction of the narrower band may be detected via the quantum
interference, and shows up as the Fano lineshape. Thus, both the
reference and the resonance channels for Fano interference are
readily identified. Since the density of states and the effective
couplings between the molecular levels and leads are tunable by
making use of the magnetic flux and gate-voltages, several ways to
control Fano lineshape are proposed, including the total flux
$\phi$, the flux difference between two sub-rings $\theta$, the
inter-dot coupling strength $t_c$, and the dot-lead coupling. In
these ways, we may realize the swap effect, the flipped tail
direction of the Fano lineshape, and the continuous
lineshape-keeping magnetic switch in this DQD system, which might
be of practical applications.

\begin{acknowledgments}
We would like to acknowledge Hui Zhai, Zuo-zi Chen and Chaoxing
Liu for helpful discussions. This work is supported by  the
Natural Science Foundation of China (Grant No. 10374056), the MOE
of China (Grant No. 2002003089), and the Program of Basic Research
Development of China (Grant No. 2001CB610508).

\end{acknowledgments}

\end{document}